\documentclass{article}

\usepackage[utf8]{inputenc}

\usepackage[margin=1.25in]{geometry}
\usepackage[labelfont=bf,font={small}]{caption}
\usepackage[strict]{changepage} 
\usepackage{graphicx}
\usepackage{epstopdf}

\usepackage{mathtools}        
\usepackage{bbm}

\usepackage{color}
\usepackage[dvipsnames]{xcolor}

\usepackage[pdftex,colorlinks=true,linkcolor=blue,citecolor=blue]{hyperref}

\usepackage{placeins}

\usepackage{authblk}

\usepackage{amsfonts}
\usepackage{physics}
\usepackage{nicefrac} 

\usepackage[numbers, sort&compress]{natbib}

\usepackage{hhline}
\usepackage{multirow}

\usepackage{soul}

    \usepackage{lineno}   
    \usepackage{amsmath}  
    \usepackage{etoolbox} 

    \newcommand*\linenomathpatch[1]{%
    \cspreto{#1}{\linenomath}%
    \cspreto{#1*}{\linenomath}%
    \csappto{end#1}{\endlinenomath}%
    \csappto{end#1*}{\endlinenomath}%
    }

    \linenomathpatch{equation}
    \linenomathpatch{gather}
    \linenomathpatch{multline}
    \linenomathpatch{align}
    \linenomathpatch{alignat}
    \linenomathpatch{flalign}

\renewcommand{\v}[1]{\ensuremath{\boldsymbol{\mathbf{#1}}}}

\begin{document}

\title{Engineering ideal helical topological networks in stanene via Zn decoration}

\author[1*]{Jennifer Coulter}
\author[2$\dag$*]{Mark R. Hirsbrunner}
\author[2]{Oleg Dubinkin}
\author[2]{Taylor L. Hughes}
\author[1]{Boris Kozinsky}

\affil[1]{\footnotesize{Harvard John A. Paulson School of Engineering and Applied Sciences, Harvard University, Cambridge, MA, 02138, USA}}
\affil[2]{\footnotesize{Department of Physics and Institute for Condensed Matter Theory, University of Illinois at Urbana-Champaign, Urbana, IL 61801, USA}}
\affil[$\dag$]{\footnotesize{To whom correspondence should be addressed: hrsbrnn2@illinois.edu}}
\affil[*]{\footnotesize{These authors contributed equally}}

\date{\today}

\maketitle


\vspace{12pt}

The xene family of topological insulators plays a key role in many proposals for topological electronic, spintronic, and valleytronic devices. These proposals rely on applying local perturbations, including electric fields and proximity magnetism, to induce topological phase transitions in xenes. However, these techniques lack control over the geometry of interfaces between topological regions, a critical aspect of engineering topological devices. We propose adatom decoration as a method for engineering atomically precise topological edge modes in xenes. Our first-principles calculations show that decorating stanene with Zn adatoms exclusively on one of two sublattices induces a topological phase transition from the quantum spin Hall (QSH) to quantum valley Hall (QVH) phase and confirm the existence of spin-valley polarized edge modes propagating at QSH/QVH interfaces. We conclude by discussing technological applications of these edge modes that are enabled by the atomic precision afforded by recent advances in adatom manipulation technology.


\vspace{12pt}


Topological edge states are protected, gapless modes that appear at the interface between regions with topologically distinct band structures~\cite{hasan_colloquium_2010, qi_topological_2011}. The topological protection and spin-momentum locking of quantum spin Hall (QSH) insulator edge modes make them promising for mesoscopic and nanoscale device applications. In the decade following the first material realization of a QSH phase, the band gaps of all known materials were too small to be of use in technological applications at room temperature, where thermal excitations mask the topological properties. However, in recent years, a large variety of two dimensional topological phases with large band gaps have been predicted and synthesized, including quantum spin Hall, quantum anomalous Hall, and quantum valley Hall (QVH) phases~\cite{ren_topological_2016}. This is an exciting development for the field of nanotechnology, as a large gap extends the viability of topological edge modes to room temperature operation~\cite{katsuragawa_room-temperature_2020}. A wide variety of electronic, spintronic, and valleytronic devices that make use of these edge modes have been theorized, ranging from ``designer'' interconnect networks~\cite{george_chiral_2012} to valley filters~\cite{jana_robust_2021}. However, many challenges remain on the path to realizing nanoscale topological devices. Here we address one of these critical challenges: reliably fabricating nanoscale interfaces between topologically distinct regions while maintaining the utility of the edge states.

There are three methods for fabricating topological edge modes: physically terminating a topological material to form an interface with the vacuum, constructing a heterostructure of topologically distinct materials, or artificially inducing spatial domain walls that act as local topological phase transitions in a single material. Despite the topological protection afforded by the bulk, manufacturing useful edge modes in these ways is challenging. Physical edges of two dimensional materials can only be milled or patterned with a limited amount of precision~\cite{mambakkam_fabrication_2020, kollipara_optical_2020, stanford_emerging_2018}, and physical edges break the translation symmetry that protects some topological edge modes\cite{xiao_valley-contrasting_2007}. Lateral heterostructures of two-dimensional materials are extremely challenging to construct, and charge transfer effects can form Schottky barriers, p-n junctions, or even misalign the band gaps resulting in a metallic device~\cite{loh_grapheneboron_2015,  li_lateral_2015, cui_chemical_2018, wang_recent_2019}. Inducing topological domain walls through electric fields, strain, and chemical functionalization is possible~\cite{Ni12, ezawa_monolayer_2015, molle_buckled_2017, krawiec_functionalization_2018}, but techniques for applying these perturbations with a high degree of local precision are in their infancy~\cite{kollipara_optical_2020, stanford_emerging_2018}.

Here we propose a novel technique to induce local topological phase transitions in stanene, a monolayer atomic film of tin atoms that hosts a large band gap QSH phase~\cite{molle_buckled_2017}, that avoids all of the above challenges and is achievable with existing technologies. Theoretical studies have demonstrated that manipulating the sublattice degree of freedom in xenes can induce a QSH to QVH phase transition~\cite{ezawa_monolayer_2015}. Because xenes form a buckled honeycomb lattice, it is possible to achieve this by applying a vertical electric field, producing a potential difference between lattice sites~\cite{ezawa_monolayer_2015}. Unfortunately, extremely strong electric fields are required to cause the transition in large gap QSH phases, because the electric potential difference along the buckling direction must be comparable to the band gap. Additionally, patterning voltage gates on two-dimensional materials to generate these electric fields is very challenging~\cite{kollipara_optical_2020, stanford_emerging_2018}. Here we explore an alternative possibility of manipulating the sublattice degree of freedom by sublattice-selective adatom decoration. Through first-principles calculations, we show that decorating one of the two sublattices of stanene with Zn adatoms generates a sufficiently large energy difference between the sublattices to induce the QSH to QVH transition. Through further large scale electronic structure calculations of decorated stanene nanoribbons, we confirm the existence of topological edge states between bare and decorated regions, and between regions decorated on opposite sublattices. This atomic-scale control of edge state interfaces will enable new atomically-defined device geometries as depicted in Fig. \ref{fig:devices}. 

A recent \emph{ab-initio} study of metallic adatoms on stanene hypothesized that Zn might form a weak van der Waals bond with stanene, rather than an ionic or covalent bond\cite{xing2017tunable}. Importantly, we will see that the weak nature of the Zn-stanene bond causes Zn-decoration to act as an ideal sublattice perturbation. Indeed the adatom decoration generates band inversions at the $\v{K}$ and $\v{K}'$ points of the Brillouin zone with no other significant modifications of the bands. Furthermore, the Zn atoms do not exchange a significant amount of charge with the stanene monolayer, resulting in closely aligned Fermi levels and band gaps between decorated and bare regions. The combined result of these effects is the ability to create ideal topological interfaces that host dissipationless edge states between decorated and bare regions of stanene. This platform is also well-suited for fabrication with existing technology because the decorated domains can be patterned at the atomic scale by manipulation of adatom positions with a scanning tunneling microscope (STM) tip. This is a well-established technique that has been used to create intricate geometries of adatoms and vacancies in many different systems and is capable of engineering arbitrary device geometries out of these topological edge modes~\cite{ko_atomicscale_2019, khajetoorians_creating_2019, fang_towards_2019}.
    
\begin{figure}
    \includegraphics[width=1.0\textwidth]{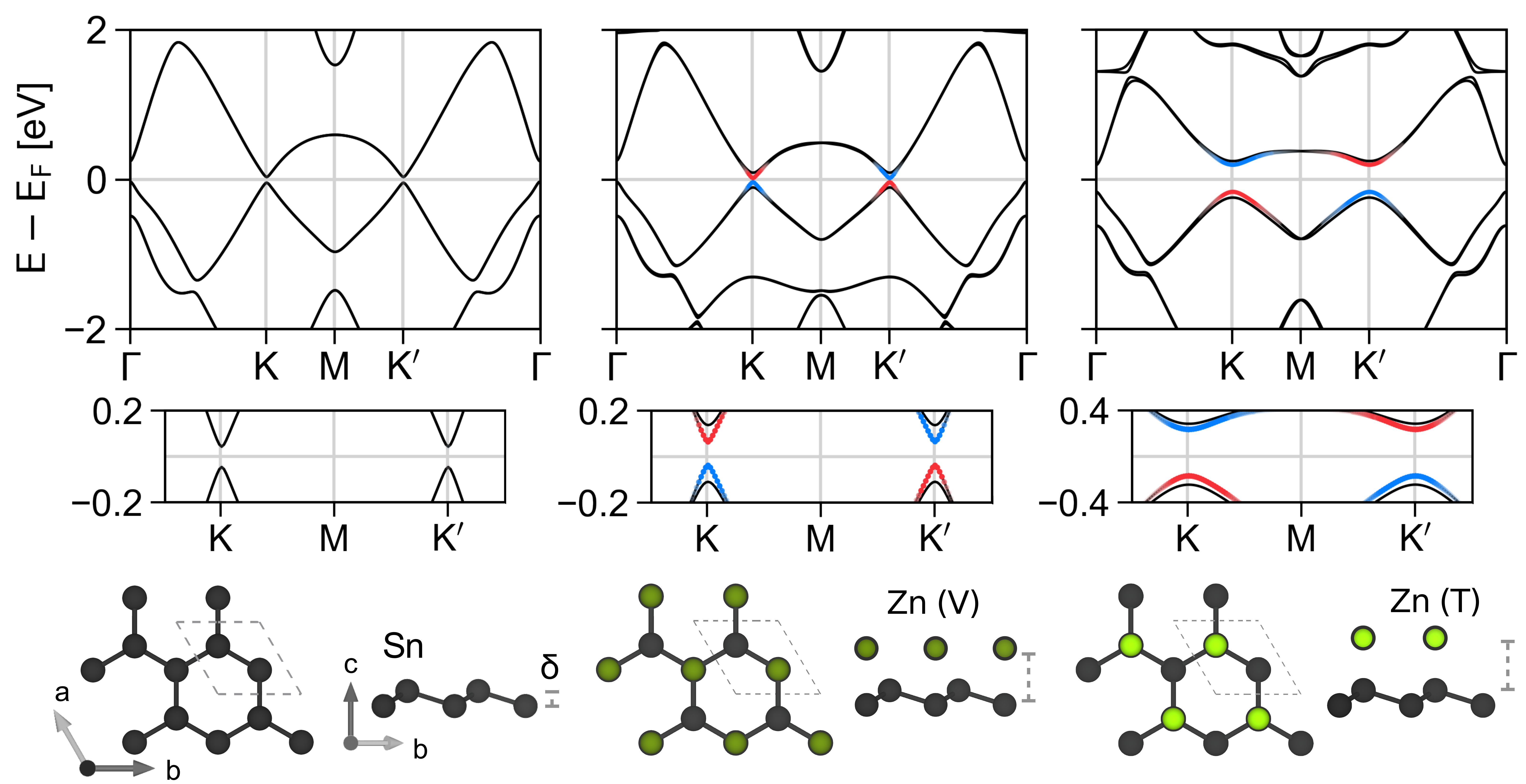}
    \caption{
    \textbf{Structure, energy bands, and Berry curvature of functionalized stanene.}
    The top row shows the calculated electronic structure of (a) bare, (b) V-decorated, and (c) T-decorated stanene. Positive and negative Berry curvature is indicated with red and blue shading, respectively. The middle row provides an enlarged view of the gaps to show the Berry curvature distribution more clearly. The bottom row displays the structures of bare, V-decorated, and T-decorated stanene and indicates the buckling height of stanene, $\delta$.
    }
    \label{fig:struc_BC}
\end{figure}
        
\section{Band Structure Calculations}
To support these claims, we performed extensive ab-initio numerical simulations of adatom-decorated stanene. For all of our simulations, we used the high-buckled, free-standing structure of stanene shown in Fig.~\ref{fig:struc_BC}, with buckling height $\delta$=0.859\AA~, and in-plane lattice parameter $a=4.676$\AA,~as found by relaxing the structure of free-standing stanene using DFT. We decorated stanene with Zn adatoms at one of each of the four structural sites for the buckled honeycomb lattice: hollow (H), bridge (B), valley (V), and top (T)~\cite{naqvi2017exploring}, and relaxed the height of the adatoms. Because the phenomena in which we are interested require sublattice site symmetry breaking, we primarily focus on the V and T adatom sites for the remainder of this work.

To determine the stability of Zn atoms in the V and T positions, we used density functional theory to calculate the adsorption energy of Zn at each site using the definition  $E_{ads} = E_{Zn+stanene} - E_{stanene} - E_{Zn}$. We found the adsorption energies to be $E_{ads}^{V} = -0.404$ eV and $E_{ads}^{T} = -0.545$ eV. To explore the possibility that the adatoms would migrate from the V and T positions, we also performed both a nudged elastic band calculation~\cite{henkelman2000climbing} to determine the most favorable pathway for adatom transport away from the V or T sites, and a dimer method calculation~\cite{henkelman1999dimer} to precisely determine the activation barriers for Zn migration. These calculations indicated that the Zn atoms on V and T sites are most likely to move to the H site, with diffusion barriers $E^{V}_{barrier}$ = 0.008 eV and $E^{T}_{barrier}$ = 0.011 eV. This is similar to other reports of barrier heights for the migration of adatoms on stanene for the V/T $\rightarrow$ H processes~\cite{mortazavi2016staneneAdatomNaLi} and indicates that the adatoms will be slow to diffuse at temperatures below $\sim$100K. These migration barriers, while limiting stability at higher temperature, would also allow the adatoms to be manipulated by STM techniques more easily. However, for cases where a higher operating temperature is desirable, we identify other candidate adatoms with higher diffusion barriers in the supplemental material.
    
Next we calculated the band structure for bare stanene and stanene decorated with Zn adatoms at the T or V sites, shown in Fig~\ref{fig:struc_BC}a-c.  Bare stanene has massive Dirac cones with negative band gaps of magnitude $E_g^{bare} = 0.073$ eV at the $\v{K}$ and $\v{K}'$ points. When decorated with Zn adatoms in either position, the degenerate Dirac cones are spin-split, resulting in a smaller $E_g^{V}=0.095$ eV gap for V decoration and a larger $E_g^{T}=0.398$ eV gap for T decoration. Crucially, we find that the decoration leaves the bands away from the Fermi level qualitatively unchanged, ensuring that the significant physical changes in the material can be captured by the topological indices near the Fermi-level. 

We repeated this analysis for stanene decorated at the T and V sites with each element in rows 2 through 5 of the periodic table. We find that nearly all elements produce bands that differ significantly from bare stanene. Additionally, many elements that do not qualitatively change the band structure of stanene end up doping the system to result in a metallic character. We provide more details of the viability of these adatom species and how they compare to Zn in the supplementary material.

\section{QSH and QVH Indicators}
Using the results of the ab-initio calculations, we generated tight-binding parameters via the maximally-localized Wannier function procedure. From the resulting Hamiltonians we calculate the Berry curvature for bare and decorated stanene~\cite{marzari2012maximally}. As shown by the coloration of the band structures in Fig.~\ref{fig:struc_BC}b and c, with red and blue representing positive and negative Berry curvature, both decorations produce equal and opposite Berry curvature concentrations at the $\v{K}$ and $\v{K}'$ points, indicating that one set of bands at each of these points was inverted by the Zn decoration. 

The origin and consequences of these band inversions are best understood via a low-energy effective model for the massive Dirac cones at the $\v{K}$ and $\v{K}'$ points in stanene~\cite{yao_spin-orbit_2007, Liu11, molle_buckled_2017}:
\begin{equation}
    \begin{split}
        H=\hbar v_F(\eta k_x\tau^x+ k_y\tau^y)+\eta\tau^z\sigma^z\lambda_{SO}+\Delta\tau^z,
    \end{split}
    \label{eqn:ham_tb}
\end{equation}
where $\tau^\alpha$ and $\sigma^\alpha$ are Pauli matrices for the sublattice and spin degrees of freedom respectively, $v_F$ is the Fermi velocity, $\eta=+1$ for $\v{K}$ and $\eta=-1$ for $\v{K}'$, and $\lambda_{SO}$ is the spin-orbit coupling strength. The final term describes a staggered potential of strength $\Delta$ between the sublattice sites generated by the adatom decoration.

In the absence of the staggered potential $\Delta$, this model describes the QSH phase realized by bare stanene. Because the Berry curvature distribution is concentrated around the $\v{K}$ and $\v{K}'$ points, and the $z$-component of the spin is conserved, we can define spin-valley resolved Chern numbers, which are protected by time-reversal symmetry and spin-conservation, by integrating the Berry curvature of a particular spin around $\v{K}$ or $\v{K}'$~\cite{ezawa_monolayer_2015}. We obtain $C_{K\uparrow}=C_{K'\uparrow}=\pm1/2$ and $C_{K\downarrow}=C_{K'\downarrow}=\mp1/2$, with the signs dependent on the sign of $\lambda_{SO}$. In terms of these spin-valley resolved indices, the total Chern number $C \in \mathbb{Z}$ and the spin Chern number $C_s \in \mathbb{Z}_2$ are
\begin{equation}
    \begin{aligned}
        C &= C_{K\uparrow} + C_{K\downarrow} + C_{K'\uparrow} + C_{K'\downarrow} = 0 \\
        C_s &= \frac{1}{2}(C_{K\uparrow} - C_{K\downarrow} + C_{K'\uparrow} - C_{K'\downarrow}) = 1 \text{ mod } 2.
    \end{aligned}
\end{equation}
According to the bulk-boundary correspondence, we expect that interfaces between regions with different spin Chern numbers host gapless helical modes that carry a spin current\cite{kanemele05, Bernevig06, hasan_colloquium_2010, qi_topological_2011}. Pairs of helical modes are not protected by time-reversal symmetry, so the spin Chern number is defined modulo $2$, $C_s\in\mathbbm{Z}_2$, and interfaces either have zero or one pair of stable helical modes.

Now when we consider the adatom decoration we find that a sufficiently large positive (negative) sublattice potential $\Delta$ induces a spin-valley resolved band inversion and changes the signs of $C_{K,\uparrow}$ and $C_{K',\downarrow}$ ($C_{K,\downarrow}$ and $C_{K',\uparrow}$)~\cite{Ni12, ezawa_monolayer_2015}. The resulting Chern and spin Chern numbers both vanish, but we can instead assign a translation symmetry protected \emph{momentum} vector charge to each valley.
This charge is equal to the vector describing the position of the valley in momentum space and defines the valley vector index: $\vec{C}_{v}=\hbar\vec{K}(C_{K}-C_{K'})=2\hbar\vec{K}$.
Systems with a non-vanishing $\vec{C}_v$ are called quantum valley Hall (QVH) insulators~\cite{xiao_valley-contrasting_2007, ren_topological_2016}. 

Interestingly, a change in the QVH index across an interface is accompanied by a translation symmetry protected current~\cite{Xiao07} carrying \emph{lattice momentum} along the interface. 
The amount of momentum transported along the edge is characterized by a scalar quantity $C_v$, equal to the projection of $\vec{C}_v$ onto the unit vector $\vec{\tau}$ tangential to the interface.
In particular, a straight open edge satisfying $\vec{\tau} \cdot \vec{C}_v = 0$ projects the valleys on top of each other, resulting in a trivial edge as indicated by the vanishing scalar index $C_v=0$. In contrast, when $\vec{\tau}$ is parallel to $\vec{C}_v$, we obtain $C_v=2\hbar|\vec{K}|$. It is customary to drop the factor of momentum $\hbar|\vec{K}|$ from the definition of $C_v$ altogether and work with the dimensionless valley index $C_{v}=C_{K}-C_{K'}=\pm2$. We list the values of the Chern numbers and valley index realized by bare and decorated stanene in Table~\ref{tab:indices}.

\begin{table}[t]
    \centering
    \def\arraystretch{1.25}
    \begin{tabular}{| c | c c c | c |}
        \hline
        Phase & $C$ & $C_{s}$ & $C_{v}$ & Decoration pattern \\
        \hline
        \hline
        
        \multirow{2}{*}{QVH} & 0 & 0 & 2 & Zn at V  \\
                              \cline{2-5}
                              & 0 & 0 & $-2$ & Zn at T \\
        \hline
        QSH & 0 & 1 & 0 & No adatom decoration \\
        \hline
    \end{tabular}
  \caption{The Chern number, $C$, spin-Chern number, $C_s$, and valley-Chern number, $C_{v}$, for each topologically nontrivial phase realized by Hamiltonian (\ref{eqn:ham_tb}), along with the corresponding adatom decoration patterns.}
  \label{tab:indices}
\end{table}
    
\begin{figure}
    \centering
    \includegraphics[width=1.0\textwidth]{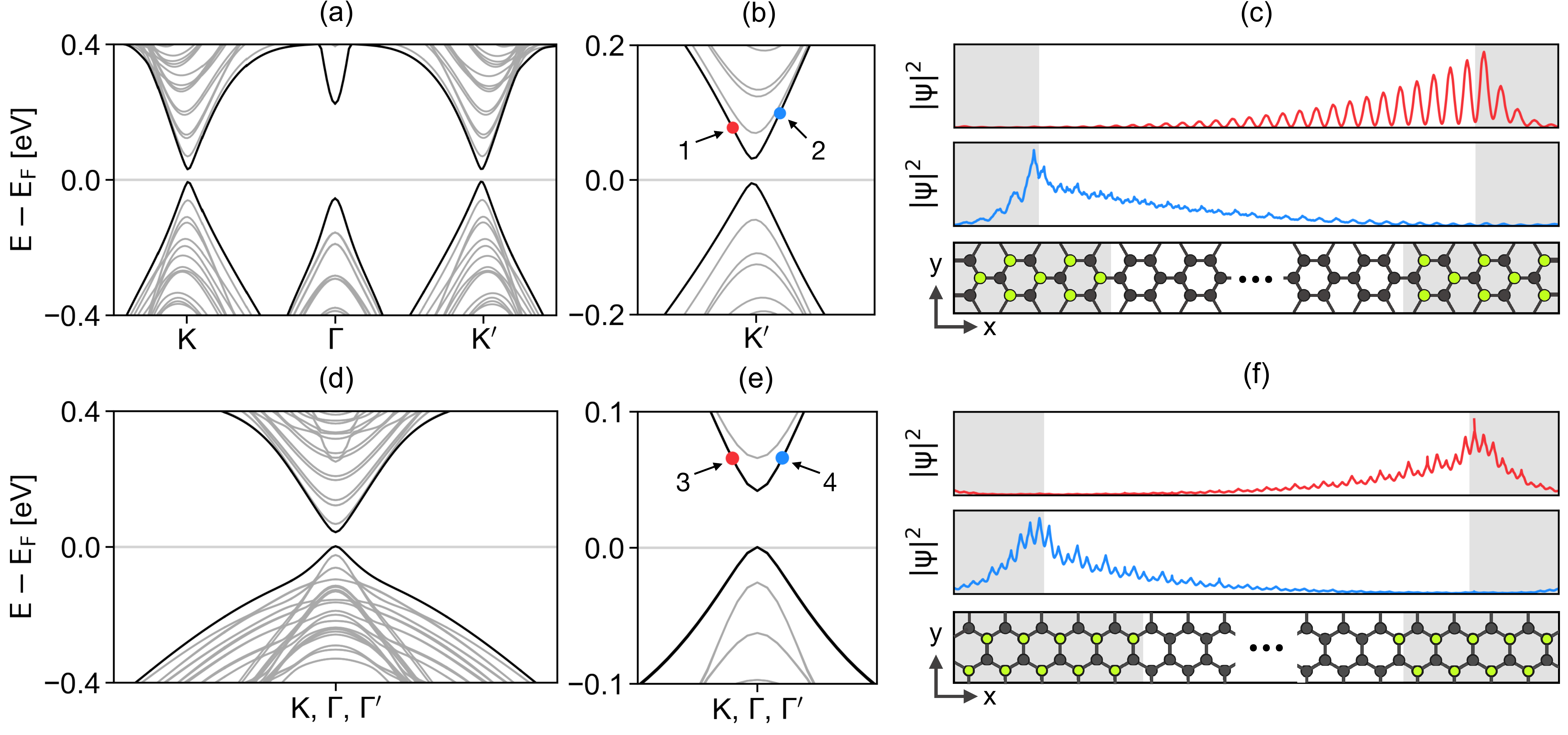}
    \caption{
    \textbf{Interfacial edge modes from the QSH-QVH structure.}
    The (a, b, d, e) band structure and (c, f) representative edge state probability distributions for the QSH-QVH ribbon for zigzag (top) and armchair (bottom) orientations. In (a, b, d, e), the thin gray and thick black lines represent the bulk and edge states, respectively. (b, e) zoom in on the near-gap region at the $\v{K}'$ point. The red dots marks a spin-up edge state propagating down the right interface and the blue dots marks a spin-up edge state propagating up the left interface. The line plots in (c) and (f) represent the probability density of the edge states indicated by the red and blue dots integrated over the plane perpendicular to the width of the ribbon. The shapes of the probability density plots for the zigzag interfaces differ because the adatom decoration makes the two interfaces asymmetric. An in-plane view of the interface structure, with colored atoms corresponding to those in Fig.~\ref{fig:struc_BC} is shown in the bottom row of (c) and (f). The locking of the spin and valley degrees of freedom at each interface is the hallmark of the QSH-QVH edge state.
    }
    \label{fig:qsh_qvh}
\end{figure}

\section{First-Principles Interface Calculations}
With this understanding of the bulk properties of decorated and bare stanene we can now consider interfaces between different spatial domains. Three interfaces can be constructed from the three phases realized by bare and decorated stanene: two distinct QSH-QVH interfaces where the spin Chern number changes from 1 to 0 and the valley Chern number changes from 0 to $\pm2$, and a QVH-QVH interface where the spin Chern number is zero on both sides while the valley Chern number changes from $-2$ to $2$. As discussed above, the QVH-QVH interface is sensitive to the orientation of the interface relative to the valley separation, $\mathbf{K}-\mathbf{K}'$, so we consider only ``zigzag'' QVH-QVH interfaces for which the edge is perpendicular to the valley separation. The QSH-QVH interfaces are insensitive to the edge orientation because the change in $C_s$ does not depend on the valley degree of freedom, so we consider both zigzag and armchair interfaces for this case. 

Now let us consider the possible interface states. At QVH-QVH interfaces, the valley Chern number changes from $\mp2$ to $\pm2$, indicating that four gapless edge modes will appear. Each valley hosts two chiral modes, with the chirality determined by the valley such that a net momentum current will flow along the interface. At QSH-QVH interfaces, the spin Chern number changes by one, and the valley Chern number changes by two, producing a pair of oppositely propagating spin-valley polarized modes. The modes in each valley are of opposite spin and opposite chirality, resulting in both spin and momentum currents at the interface.

\begin{figure}
    \centering
    \includegraphics[width=1.0\textwidth]{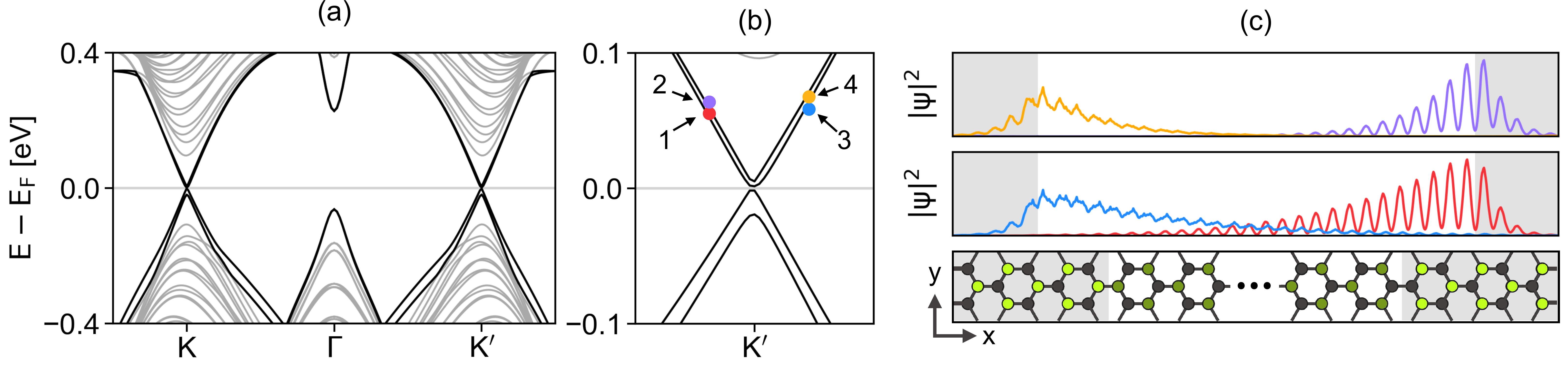}
    \caption{
    \textbf{Interfacial edge modes from the QVH-QVH structure.}
    The (a, b) band structure and (c) representative edge state probability densities of the QVH-QVH interface with a zigzag orientation. In (a, b), the thin gray and thick black lines represent the bulk and edge states, respectively. (b) Zoom in on the near-gap region at the $\v{K}'$ point. The red and purple dots mark states propagating along the right interface and the blue and yellow dots mark edge states propagating along the left interface. The line plots in (c) represent the probability density integrated over the plane perpendicular to the width of the ribbon. The shapes of the probability density plots for each interface differ because the adatom decoration makes the two interfaces asymmetric. An in-plane view of the structure, with colored atoms corresponding to those in Fig.~\ref{fig:struc_BC}, is shown in the bottom row of (c). Each valley contributes two unpolarized edge modes to each interface, as the indicated by the change of the valley Chern number by four across each interface.
    }
    \label{fig:qvh_qvh}
\end{figure}

To determine the characteristics of these interface modes in decorated stanene, we performed large-scale first principles electronic structure simulations of stanene nanoribbons decorated to produce QVH-QSH and QVH-QVH interfaces. The translation-invariant direction points in the $\vec{b}$ and $\frac{1}{2}\vec{a} + \vec{b}$ directions to realize zigzag and armchair interfaces, respectively. The unit vectors $\vec{a}$ and $\vec{b}$ point along the primitive lattice vectors, as shown in Fig.~\ref{fig:struc_BC}. To create topological interfaces, we selectively decorated domains in the transverse direction (the x-axis in Fig.~\ref{fig:qsh_qvh}c, \ref{fig:qsh_qvh}f, and \ref{fig:qvh_qvh}c). We used periodic boundary conditions in the transverse direction to eliminate spurious interfaces with the vacuum, forming two topological interfaces per ribbon. All zigzag ribbons were 145.67~\AA ~wide and the armchair ribbon was 149.64~\AA  ~wide.

 To make the most of finite computational resources, the relative sizes of the decoration domains were chosen to minimize wavefunction overlap between the exponentially decaying interface states. Accordingly, the zigzag QSH-QVH ribbon is T-decorated on 10 unit cells and bare on 26 unit cells, the armchair QSH-QVH ribbon is T-decorated on 10 unit cells and bare on 22 unit cells, and the QVH-QVH ribbon is T-decorated on 10 unit cells and V-decorated on 26 unit cells. We note that the overlap of the interface wavefunctions in the bulk of the ribbon leads to undesired gaps in the interface spectrum produced by finite-size effects. We show in the supplementary material that these interface spectrum gaps vanish for sufficiently wide ribbons, and we find that the interface states decay exponentially into the insulating bulk with a decay length roughly determined by the ratio of the Fermi velocity to the bulk gap. 

The resulting band structure and interface wavefunction plots are shown in Fig.~\ref{fig:qsh_qvh} for the QSH-QVH zigzag and armchair ribbons. Each interface in the zigzag ribbon hosts a helical pair of spin-valley locked modes that produce both equilibrium spin and momentum currents on the interface. Each interface of the armchair ribbon also hosts a helical pair of modes, but since the valleys in this case are projected to the $\Gamma$ point of the Brillouin zone the interfaces only carry spin current, not momentum current. The configurations of edge states we find agree with the $\v{k}\cdot\v{p}$ model predictions of the previous section. As mentioned above, both ribbons have a small gap in the interface state spectrum, $E_g \approx 0.03$ eV, that originates from the overlap and hybridization of the  wavefunctions on the two interfaces and would vanish for a larger system. The decay lengths of the interface states in the zigzag ribbon are $\lambda_T=5.95$ \AA{} and $\lambda_0=32.4$ \AA{} in the T-decorated and bare regions, respectively. The decay lengths in the armchair ribbon are $\lambda_T=6.55$ \AA{} and $\lambda_0=35.7$ \AA{}.

The results for the QVH-QVH ribbon are shown in the same format as the QSH-QVH ribbon in Fig.~\ref{fig:qvh_qvh}. Each interface hosts two chiral modes from each valley, with the valleys contributing modes of opposite chirality. This leads to the equilibrium edge momentum current predicted above. In this case the decay lengths of the edge states are $\lambda_T=5.68$ \AA{} and $\lambda_V=23.8$ \AA{} in the T- and V-decorated regions, respectively. The gaps in the interface state spectrum resulting from wavefunction overlap are $E_g \approx 0.02$ eV and and $0.006$ eV. We report two gaps here because there are two sets of interface states at each valley for this ribbon. For interfaces with a finite projection onto the valley separation direction, the momentum carried by the edge is reduced. In the extreme case of an armchair interface, the valleys exactly overlap, the edge carries no momentum current, and any local perturbation can gap out the interface states. 
    
\section{Technological Applications}
\begin{figure}
    \centering
    \includegraphics[width=0.75\textwidth]{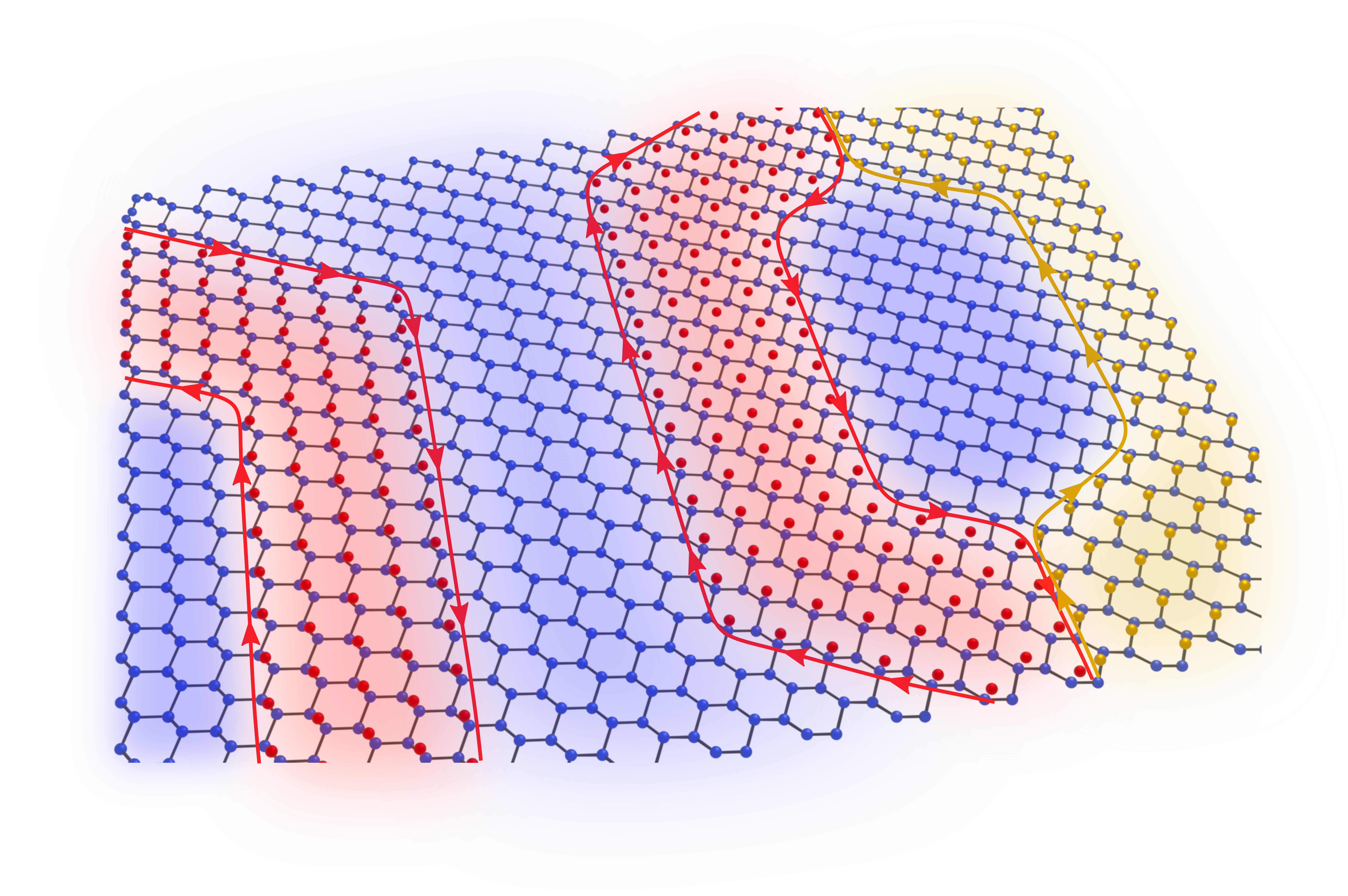}
    \caption{
    \textbf{Schematic drawing of example devices.}
    A schematic showing two possible devices constructed via adatom decoration of stanene. The blue spheres are Sn tin atoms and the red and yellow spheres are Zn adatoms located at T and V sites, respectively. The blue, red, and yellow shading is a guide to the eye, indicating the regions that are bare, T-decorated, and V-decorated, respectively. The red and yellow arrows indicate quantum spin Hall edge modes, the color determined by the decoration site of the quantum valley Hall region. The left side of the image shows two densely-packed chiral interconnects constructed by decorating a thin region of stanene with Zn adatoms. The right side of the image shows a Mach-Zehnder interferometer built out of the edge modes of two adjacent T- and V-decorated regions.
    }
    \label{fig:devices}
\end{figure}

The above calculations demonstrate that decorating stanene with Zn adatoms presents a uniquely promising platform for technological applications. The topological domains can be patterned with a high degree of control by manipulating the adatom positions with an STM tip, permitting the fabrication of many topological devices, two of which are depicted schematically in Fig.~\ref{fig:devices}. Furthermore, the interface states residing at domain walls are localized on the scale of tens of nanometers, which permits extremely dense packing of features. One of the first proposals for an application of topological edge modes was designer interconnect networks, which are a possible solution to the ``interconnect bottleneck'', wherein scattering and parasitic capacitance in interconnects leads to signal delays that prohibit further miniaturization of semiconductor devices~\cite{george_chiral_2012}. The minimum metal pitch, or center to center distance between interconnects, of the current semiconductor manufacturing technology node is 24 nm to 36 nm~\cite{ITRS}. At this scale, grain boundary and defect scattering leads to large resistances that inhibit the performance of traditional copper interconnects~\cite{graham_resistivity_2010}. Considering the edge state decay lengths obtained in our simulations, the minimum pitch that could be achieved with Zn-decorated stanene interconnects is also on the order of tens of nanometers. However, the topological protection of the interface modes eliminates the issue of scattering, drastically improving performance with no compromise on feature size.

The interface modes of Zn-decorated stanene also have many applications beyond the world of conventional electronics. The fields of spin- and valley-tronics attempt to process information by exploiting the spin and valley degrees of freedom, rather than the charge degree of freedom~\cite{bader_spintronics_2010, vitale_valleytronics_2018}. Topological interface modes are useful for engineering spin- and valleytronic devices such as waveguides, splitters, valves, and filters~\cite{li_valley_2018, ezawa_topological_2013, xu_manipulating_2017,yang_topological_2020, qiao_spin-polarized_2011, jana_robust_2021}, and STM manipulation of Zn adatoms on stanene provides an ideal platform to fabricate the precise geometries of such devices. The same is true of electron quantum optics devices, such as valley Hall beam splitters, Mach-Zehnder interferometers, and Fabry-Perot resonators~\cite{jo_quantum_2021, rickhaus_transport_2018, wei_mach-zehnder_2017}. This approach to engineering topological interface modes is also well suited to fabricating quantum computing gates out of helical edge states decorated with magnetic impurities~\cite{niyazov_coherent_2020, chen_quantum_2014}, as STM manipulation of adatoms can be used to both create the edge states \emph{and} deposit magnetic impurities.

\section{Conclusion}
We have demonstrated that sublattice-selective decoration of stanene with Zn adatoms is an excellent platform for engineering topological interface modes. Because Zn adatoms bond relatively weakly to stanene, they act as an ideal sublattice potential and induce a QSH to QVH transition in stanene. The weak nature of the bond also does not transfer significant charge to stanene and permits STM manipulation of the adatoms allowing detailed patterning of topological interfaces. Importantly, the Zn-Sn bond is also strong enough for the decoration to remain stable at liquid nitrogen temperatures. The combined result of these effects is a platform suitable for fabricating arbitrary networks of topological interface modes without any of the deleterious effects that plague existing proposals for topological devices. These ideal topological interface-state networks have applications in semiconductor devices, spintronics, valleytronics, quantum electron optics, and even quantum computing. Implementing this technique is possible with existing fabrication and STM technology and can lead to transformative advances in topological device engineering.

\section*{Methods}
    The investigation of all possible adatom species in rows two through five of the periodic table was completed using JDFTx~\cite{sundararaman2017jdftx} to take advantage of GPU functionality. These calculations were carried out with ONCV pseudopotentials~\cite{hamann2013optimized, van2018pseudodojo} using the Perdew-Burke-Enzerhof (PBE)~\cite{pbe} exchange-correlation functional, a 1090 eV plane-wave energy cutoff, a 15x15x1 $\Gamma$-centered k-mesh, and Methfessel-Paxton smearing of $\sigma = 0.0272$ eV. 
    
    The electronic structure of the decorated interface structures was determined via the Vienna Ab-initio Software Package (VASP)~\cite{vasp1,vasp2,vasp3} using the PBE functional with the projector-augmented wave (PAW)~\cite{paw_pseudopotentials} potentials provided by VASP. The calculations were performed on a 1x15x1 $\Gamma$-centered k-mesh with a plane-wave energy cutoff of 450 eV and Methfessel-Paxton smearing of $\sigma = 0.01$ eV.
    
    All calculations included spin-orbit coupling, 20 \AA{} of z-axis vacuum between periodic images, and the many-body dispersion (MBD) van der Waals correction~\cite{ambrosetti2014long}. Probability density data was visualized using the pawpyseed package~\cite{pawpyseed}.

\nolinenumbers

\bibliographystyle{naturemag}
\footnotesize{\bibliography{main.bib}}

\begin{thebibliography}{10}
\expandafter\ifx\csname url\endcsname\relax
  \def\url#1{\texttt{#1}}\fi
\expandafter\ifx\csname urlprefix\endcsname\relax\def\urlprefix{URL }\fi
\providecommand{\bibinfo}[2]{#2}
\providecommand{\eprint}[2][]{\url{#2}}

\bibitem{hasan_colloquium_2010}
\bibinfo{author}{Hasan, M.~Z.} \& \bibinfo{author}{Kane, C.~L.}
\newblock \bibinfo{title}{\textit{{Colloquium}} : {Topological} insulators}.
\newblock \emph{\bibinfo{journal}{Reviews of Modern Physics}}
  \textbf{\bibinfo{volume}{82}}, \bibinfo{pages}{3045--3067}
  (\bibinfo{year}{2010}).
\newblock \urlprefix\url{https://link.aps.org/doi/10.1103/RevModPhys.82.3045}.

\bibitem{qi_topological_2011}
\bibinfo{author}{Qi, X.-L.} \& \bibinfo{author}{Zhang, S.-C.}
\newblock \bibinfo{title}{Topological insulators and superconductors}.
\newblock \emph{\bibinfo{journal}{Reviews of Modern Physics}}
  \textbf{\bibinfo{volume}{83}}, \bibinfo{pages}{1057--1110}
  (\bibinfo{year}{2011}).
\newblock \urlprefix\url{https://link.aps.org/doi/10.1103/RevModPhys.83.1057}.

\bibitem{ren_topological_2016}
\bibinfo{author}{Ren, Y.}, \bibinfo{author}{Qiao, Z.} \& \bibinfo{author}{Niu,
  Q.}
\newblock \bibinfo{title}{Topological phases in two-dimensional materials: a
  review}.
\newblock \emph{\bibinfo{journal}{Reports on Progress in Physics}}
  \textbf{\bibinfo{volume}{79}}, \bibinfo{pages}{066501}
  (\bibinfo{year}{2016}).
\newblock
  \urlprefix\url{https://iopscience.iop.org/article/10.1088/0034-4885/79/6/066501}.

\bibitem{katsuragawa_room-temperature_2020}
\bibinfo{author}{Katsuragawa, N.} \emph{et~al.}
\newblock \bibinfo{title}{Room-temperature quantum spin {Hall} phase in
  laser-patterned few-layer {1T}'- {MoS2}}.
\newblock \emph{\bibinfo{journal}{Communications Materials}}
  \textbf{\bibinfo{volume}{1}} (\bibinfo{year}{2020}).
\newblock \urlprefix\url{https://www.nature.com/articles/s43246-020-00050-w}.

\bibitem{george_chiral_2012}
\bibinfo{author}{Zhang, X.} \& \bibinfo{author}{Zhang, S.-C.}
\newblock \bibinfo{title}{Chiral interconnects based on topological
  insulators}.
\newblock In \bibinfo{editor}{George, T.}, \bibinfo{editor}{Islam, M.~S.} \&
  \bibinfo{editor}{Dutta, A.} (eds.) \emph{\bibinfo{booktitle}{Micro- and
  {Nanotechnology} {Sensors}, {Systems}, and {Applications} {IV}}}, vol.
  \bibinfo{volume}{8373}, \bibinfo{pages}{71 -- 81} (\bibinfo{publisher}{SPIE},
  \bibinfo{year}{2012}).
\newblock \urlprefix\url{https://doi.org/10.1117/12.920325}.

\bibitem{jana_robust_2021}
\bibinfo{author}{Jana, K.} \& \bibinfo{author}{Muralidharan, B.}
\newblock \bibinfo{title}{Robust all-electrical topological valley filtering
  using monolayer {2D}-{Xenes}}.
\newblock \emph{\bibinfo{journal}{arXiv:2107.13318 [cond-mat]}}
  (\bibinfo{year}{2021}).
\newblock \urlprefix\url{http://arxiv.org/abs/2107.13318}.

\bibitem{mambakkam_fabrication_2020}
\bibinfo{author}{Mambakkam, S.~V.} \& \bibinfo{author}{Law, S.}
\newblock \bibinfo{title}{Fabrication of topological insulator nanostructures}.
\newblock \emph{\bibinfo{journal}{Journal of Vacuum Science \& Technology B}}
  \textbf{\bibinfo{volume}{38}}, \bibinfo{pages}{055001}
  (\bibinfo{year}{2020}).
\newblock \urlprefix\url{http://avs.scitation.org/doi/10.1116/6.0000341}.

\bibitem{kollipara_optical_2020}
\bibinfo{author}{Kollipara, P.~S.}, \bibinfo{author}{Li, J.} \&
  \bibinfo{author}{Zheng, Y.}
\newblock \bibinfo{title}{Optical {Patterning} of {Two}-{Dimensional}
  {Materials}}.
\newblock \emph{\bibinfo{journal}{Research}} \textbf{\bibinfo{volume}{2020}}
  (\bibinfo{year}{2020}).
\newblock
  \urlprefix\url{https://spj.sciencemag.org/journals/research/2020/6581250/}.

\bibitem{stanford_emerging_2018}
\bibinfo{author}{Stanford, M.~G.}, \bibinfo{author}{Rack, P.~D.} \&
  \bibinfo{author}{Jariwala, D.}
\newblock \bibinfo{title}{Emerging nanofabrication and quantum confinement
  techniques for {2D} materials beyond graphene}.
\newblock \emph{\bibinfo{journal}{npj 2D Materials and Applications}}
  \textbf{\bibinfo{volume}{2}}, \bibinfo{pages}{1--15} (\bibinfo{year}{2018}).
\newblock \urlprefix\url{https://www.nature.com/articles/s41699-018-0065-3}.

\bibitem{xiao_valley-contrasting_2007}
\bibinfo{author}{Xiao, D.}, \bibinfo{author}{Yao, W.} \& \bibinfo{author}{Niu,
  Q.}
\newblock \bibinfo{title}{Valley-{Contrasting} {Physics} in {Graphene}:
  {Magnetic} {Moment} and {Topological} {Transport}}.
\newblock \emph{\bibinfo{journal}{Physical Review Letters}}
  \textbf{\bibinfo{volume}{99}}, \bibinfo{pages}{236809}
  (\bibinfo{year}{2007}).
\newblock
  \urlprefix\url{https://link.aps.org/doi/10.1103/PhysRevLett.99.236809}.
\newblock \bibinfo{note}{Publisher: American Physical Society}.

\bibitem{loh_grapheneboron_2015}
\bibinfo{author}{Loh, G.~C.} \& \bibinfo{author}{Pandey, R.}
\newblock \bibinfo{title}{A graphene–boron nitride lateral heterostructure
  – a first-principles study of its growth, electronic properties, and
  chemical topology}.
\newblock \emph{\bibinfo{journal}{Journal of Materials Chemistry C}}
  \textbf{\bibinfo{volume}{3}}, \bibinfo{pages}{5918--5932}
  (\bibinfo{year}{2015}).
\newblock \urlprefix\url{http://xlink.rsc.org/?DOI=C5TC00539F}.

\bibitem{li_lateral_2015}
\bibinfo{author}{Li, Y.} \emph{et~al.}
\newblock \bibinfo{title}{Lateral and {Vertical} {Two}-{Dimensional} {Layered}
  {Topological} {Insulator} {Heterostructures}}.
\newblock \emph{\bibinfo{journal}{ACS Nano}} \textbf{\bibinfo{volume}{9}},
  \bibinfo{pages}{10916--10921} (\bibinfo{year}{2015}).
\newblock \urlprefix\url{https://pubs.acs.org/doi/10.1021/acsnano.5b04068}.

\bibitem{cui_chemical_2018}
\bibinfo{author}{Cui, Y.}, \bibinfo{author}{Li, B.}, \bibinfo{author}{Li, J.}
  \& \bibinfo{author}{Wei, Z.}
\newblock \bibinfo{title}{Chemical vapor deposition growth of two-dimensional
  heterojunctions}.
\newblock \emph{\bibinfo{journal}{Science China Physics, Mechanics \&
  Astronomy}} \textbf{\bibinfo{volume}{61}}, \bibinfo{pages}{016801}
  (\bibinfo{year}{2018}).
\newblock \urlprefix\url{http://link.springer.com/10.1007/s11433-017-9105-x}.

\bibitem{wang_recent_2019}
\bibinfo{author}{Wang, J.}, \bibinfo{author}{Li, Z.}, \bibinfo{author}{Chen,
  H.}, \bibinfo{author}{Deng, G.} \& \bibinfo{author}{Niu, X.}
\newblock \bibinfo{title}{Recent {Advances} in {2D} {Lateral}
  {Heterostructures}}.
\newblock \emph{\bibinfo{journal}{Nano-Micro Letters}}
  \textbf{\bibinfo{volume}{11}}, \bibinfo{pages}{48} (\bibinfo{year}{2019}).
\newblock \urlprefix\url{http://link.springer.com/10.1007/s40820-019-0276-y}.

\bibitem{Ni12}
\bibinfo{author}{Ni, Z.} \emph{et~al.}
\newblock \bibinfo{title}{Tunable bandgap in silicene and germanene}.
\newblock \emph{\bibinfo{journal}{Nano Letters}} \textbf{\bibinfo{volume}{12}},
  \bibinfo{pages}{113--118} (\bibinfo{year}{2012}).
\newblock \urlprefix\url{https://doi.org/10.1021/nl203065e}.

\bibitem{ezawa_monolayer_2015}
\bibinfo{author}{Ezawa, M.}
\newblock \bibinfo{title}{Monolayer {Topological} {Insulators}: {Silicene},
  {Germanene}, and {Stanene}}.
\newblock \emph{\bibinfo{journal}{Journal of the Physical Society of Japan}}
  \textbf{\bibinfo{volume}{84}}, \bibinfo{pages}{121003}
  (\bibinfo{year}{2015}).
\newblock \urlprefix\url{http://journals.jps.jp/doi/10.7566/JPSJ.84.121003}.

\bibitem{molle_buckled_2017}
\bibinfo{author}{Molle, A.} \emph{et~al.}
\newblock \bibinfo{title}{Buckled two-dimensional {Xene} sheets}.
\newblock \emph{\bibinfo{journal}{Nature Materials}}
  \textbf{\bibinfo{volume}{16}}, \bibinfo{pages}{163--169}
  (\bibinfo{year}{2017}).
\newblock
  \urlprefix\url{https://www-nature-com.proxy2.library.illinois.edu/articles/nmat4802}.

\bibitem{krawiec_functionalization_2018}
\bibinfo{author}{Krawiec, M.}
\newblock \bibinfo{title}{Functionalization of group-14 two-dimensional
  materials}.
\newblock \emph{\bibinfo{journal}{Journal of Physics: Condensed Matter}}
  \textbf{\bibinfo{volume}{30}}, \bibinfo{pages}{233003}
  (\bibinfo{year}{2018}).
\newblock
  \urlprefix\url{https://iopscience.iop.org/article/10.1088/1361-648X/aac149}.

\bibitem{xing2017tunable}
\bibinfo{author}{Xing, D.-X.} \emph{et~al.}
\newblock \bibinfo{title}{Tunable electronic and magnetic properties in stanene
  by 3d transition metal atoms absorption}.
\newblock \emph{\bibinfo{journal}{Superlattices and Microstructures}}
  \textbf{\bibinfo{volume}{103}}, \bibinfo{pages}{139--144}
  (\bibinfo{year}{2017}).
\newblock \urlprefix\url{https://doi.org/10.1016/j.spmi.2017.01.033}.

\bibitem{ko_atomicscale_2019}
\bibinfo{author}{Ko, W.}, \bibinfo{author}{Ma, C.}, \bibinfo{author}{Nguyen,
  G.~D.}, \bibinfo{author}{Kolmer, M.} \& \bibinfo{author}{Li, A.}
\newblock \bibinfo{title}{Atomic‐{Scale} {Manipulation} and {In} {Situ}
  {Characterization} with {Scanning} {Tunneling} {Microscopy}}.
\newblock \emph{\bibinfo{journal}{Advanced Functional Materials}}
  \textbf{\bibinfo{volume}{29}}, \bibinfo{pages}{1903770}
  (\bibinfo{year}{2019}).
\newblock
  \urlprefix\url{https://onlinelibrary.wiley.com/doi/10.1002/adfm.201903770}.

\bibitem{khajetoorians_creating_2019}
\bibinfo{author}{Khajetoorians, A.~A.}, \bibinfo{author}{Wegner, D.},
  \bibinfo{author}{Otte, A.~F.} \& \bibinfo{author}{Swart, I.}
\newblock \bibinfo{title}{Creating designer quantum states of matter
  atom-by-atom}.
\newblock \emph{\bibinfo{journal}{Nature Reviews Physics}}
  \textbf{\bibinfo{volume}{1}}, \bibinfo{pages}{703--715}
  (\bibinfo{year}{2019}).
\newblock \urlprefix\url{https://www.nature.com/articles/s42254-019-0108-5}.

\bibitem{fang_towards_2019}
\bibinfo{author}{Fang, F.} \emph{et~al.}
\newblock \bibinfo{title}{Towards atomic and close-to-atomic scale
  manufacturing}.
\newblock \emph{\bibinfo{journal}{International Journal of Extreme
  Manufacturing}} \textbf{\bibinfo{volume}{1}}, \bibinfo{pages}{012001}
  (\bibinfo{year}{2019}).
\newblock
  \urlprefix\url{https://iopscience.iop.org/article/10.1088/2631-7990/ab0dfc}.

\bibitem{naqvi2017exploring}
\bibinfo{author}{Naqvi, S.~R.}, \bibinfo{author}{Hussain, T.},
  \bibinfo{author}{Luo, W.} \& \bibinfo{author}{Ahuja, R.}
\newblock \bibinfo{title}{Exploring doping characteristics of various adatoms
  on single-layer stanene}.
\newblock \emph{\bibinfo{journal}{The Journal of Physical Chemistry C}}
  \textbf{\bibinfo{volume}{121}}, \bibinfo{pages}{7667--7676}
  (\bibinfo{year}{2017}).
\newblock \urlprefix\url{https://doi.org/10.1021/acs.jpcc.7b00468}.

\bibitem{henkelman2000climbing}
\bibinfo{author}{Henkelman, G.}, \bibinfo{author}{Uberuaga, B.~P.} \&
  \bibinfo{author}{J{\'o}nsson, H.}
\newblock \bibinfo{title}{A climbing image nudged elastic band method for
  finding saddle points and minimum energy paths}.
\newblock \emph{\bibinfo{journal}{The Journal of Chemical Physics}}
  \textbf{\bibinfo{volume}{113}}, \bibinfo{pages}{9901--9904}
  (\bibinfo{year}{2000}).
\newblock \urlprefix\url{https://doi.org/10.1063/1.1329672}.

\bibitem{henkelman1999dimer}
\bibinfo{author}{Henkelman, G.} \& \bibinfo{author}{J{\'o}nsson, H.}
\newblock \bibinfo{title}{A dimer method for finding saddle points on high
  dimensional potential surfaces using only first derivatives}.
\newblock \emph{\bibinfo{journal}{The Journal of Chemical Physics}}
  \textbf{\bibinfo{volume}{111}}, \bibinfo{pages}{7010--7022}
  (\bibinfo{year}{1999}).
\newblock \urlprefix\url{https://doi.org/10.1063/1.480097}.

\bibitem{mortazavi2016staneneAdatomNaLi}
\bibinfo{author}{Mortazavi, B.}, \bibinfo{author}{Dianat, A.},
  \bibinfo{author}{Cuniberti, G.} \& \bibinfo{author}{Rabczuk, T.}
\newblock \bibinfo{title}{Application of silicene, germanene and stanene for
  {Na} or {Li} ion storage: A theoretical investigation}.
\newblock \emph{\bibinfo{journal}{Electrochimica Acta}}
  \textbf{\bibinfo{volume}{213}}, \bibinfo{pages}{865--870}
  (\bibinfo{year}{2016}).
\newblock \urlprefix\url{https://doi.org/10.1016/j.electacta.2016.08.027}.

\bibitem{marzari2012maximally}
\bibinfo{author}{Marzari, N.}, \bibinfo{author}{Mostofi, A.~A.},
  \bibinfo{author}{Yates, J.~R.}, \bibinfo{author}{Souza, I.} \&
  \bibinfo{author}{Vanderbilt, D.}
\newblock \bibinfo{title}{Maximally localized wannier functions: Theory and
  applications}.
\newblock \emph{\bibinfo{journal}{Reviews of Modern Physics}}
  \textbf{\bibinfo{volume}{84}}, \bibinfo{pages}{1419} (\bibinfo{year}{2012}).
\newblock \urlprefix\url{https://doi.org/10.1103/RevModPhys.84.1419}.

\bibitem{yao_spin-orbit_2007}
\bibinfo{author}{Yao, Y.}, \bibinfo{author}{Ye, F.}, \bibinfo{author}{Qi,
  X.-L.}, \bibinfo{author}{Zhang, S.-C.} \& \bibinfo{author}{Fang, Z.}
\newblock \bibinfo{title}{Spin-orbit gap of graphene: {First}-principles
  calculations}.
\newblock \emph{\bibinfo{journal}{Physical Review B}}
  \textbf{\bibinfo{volume}{75}}, \bibinfo{pages}{041401}
  (\bibinfo{year}{2007}).
\newblock \urlprefix\url{https://link.aps.org/doi/10.1103/PhysRevB.75.041401}.

\bibitem{Liu11}
\bibinfo{author}{Liu, C.-C.}, \bibinfo{author}{Feng, W.} \&
  \bibinfo{author}{Yao, Y.}
\newblock \bibinfo{title}{Quantum spin {Hall} effect in silicene and
  two-dimensional germanium}.
\newblock \emph{\bibinfo{journal}{Phys. Rev. Lett.}}
  \textbf{\bibinfo{volume}{107}}, \bibinfo{pages}{076802}
  (\bibinfo{year}{2011}).
\newblock
  \urlprefix\url{https://link.aps.org/doi/10.1103/PhysRevLett.107.076802}.

\bibitem{kanemele05}
\bibinfo{author}{Kane, C.~L.} \& \bibinfo{author}{Mele, E.~J.}
\newblock \bibinfo{title}{${Z}_{2}$ topological order and the quantum spin
  {Hall} effect}.
\newblock \emph{\bibinfo{journal}{Phys. Rev. Lett.}}
  \textbf{\bibinfo{volume}{95}}, \bibinfo{pages}{146802}
  (\bibinfo{year}{2005}).
\newblock
  \urlprefix\url{https://link.aps.org/doi/10.1103/PhysRevLett.95.146802}.

\bibitem{Bernevig06}
\bibinfo{author}{Bernevig, B.~A.} \& \bibinfo{author}{Zhang, S.-C.}
\newblock \bibinfo{title}{Quantum spin {Hall} effect}.
\newblock \emph{\bibinfo{journal}{Phys. Rev. Lett.}}
  \textbf{\bibinfo{volume}{96}}, \bibinfo{pages}{106802}
  (\bibinfo{year}{2006}).
\newblock
  \urlprefix\url{https://link.aps.org/doi/10.1103/PhysRevLett.96.106802}.

\bibitem{Xiao07}
\bibinfo{author}{Xiao, D.}, \bibinfo{author}{Yao, W.} \& \bibinfo{author}{Niu,
  Q.}
\newblock \bibinfo{title}{Valley-contrasting physics in graphene: Magnetic
  moment and topological transport}.
\newblock \emph{\bibinfo{journal}{Phys. Rev. Lett.}}
  \textbf{\bibinfo{volume}{99}}, \bibinfo{pages}{236809}
  (\bibinfo{year}{2007}).
\newblock
  \urlprefix\url{https://link.aps.org/doi/10.1103/PhysRevLett.99.236809}.

\bibitem{ITRS}
\bibinfo{title}{International technology roadmap for semiconductors ({ITRS})}
  (\bibinfo{year}{2015}).
\newblock \urlprefix\url{http://www.itrs2.net/}.

\bibitem{graham_resistivity_2010}
\bibinfo{author}{Graham, R.~L.} \emph{et~al.}
\newblock \bibinfo{title}{Resistivity dominated by surface scattering in sub-50
  nm {Cu} wires}.
\newblock \emph{\bibinfo{journal}{Applied Physics Letters}}
  \textbf{\bibinfo{volume}{96}}, \bibinfo{pages}{042116}
  (\bibinfo{year}{2010}).
\newblock \urlprefix\url{https://aip.scitation.org/doi/full/10.1063/1.3292022}.

\bibitem{bader_spintronics_2010}
\bibinfo{author}{Bader, S.} \& \bibinfo{author}{Parkin, S.}
\newblock \bibinfo{title}{Spintronics}.
\newblock \emph{\bibinfo{journal}{Annual Review of Condensed Matter Physics}}
  \textbf{\bibinfo{volume}{1}}, \bibinfo{pages}{71--88} (\bibinfo{year}{2010}).
\newblock
  \urlprefix\url{https://www.annualreviews.org/doi/10.1146/annurev-conmatphys-070909-104123}.

\bibitem{vitale_valleytronics_2018}
\bibinfo{author}{Vitale, S.~A.} \emph{et~al.}
\newblock \bibinfo{title}{Valleytronics: {Opportunities}, {Challenges}, and
  {Paths} {Forward}}.
\newblock \emph{\bibinfo{journal}{Small}} \textbf{\bibinfo{volume}{14}},
  \bibinfo{pages}{1801483} (\bibinfo{year}{2018}).
\newblock
  \urlprefix\url{https://onlinelibrary.wiley.com/doi/abs/10.1002/smll.201801483}.

\bibitem{li_valley_2018}
\bibinfo{author}{Li, J.} \emph{et~al.}
\newblock \bibinfo{title}{A valley valve and electron beam splitter}.
\newblock \emph{\bibinfo{journal}{Science}} \textbf{\bibinfo{volume}{362}},
  \bibinfo{pages}{1149--1152} (\bibinfo{year}{2018}).
\newblock
  \urlprefix\url{https://www.science.org/lookup/doi/10.1126/science.aao5989}.

\bibitem{ezawa_topological_2013}
\bibinfo{author}{Ezawa, M.}
\newblock \bibinfo{title}{Topological {Kirchhoff} law and bulk-edge
  correspondence for valley {Chern} and spin-valley {Chern} numbers}.
\newblock \emph{\bibinfo{journal}{Physical Review B}}
  \textbf{\bibinfo{volume}{88}}, \bibinfo{pages}{161406}
  (\bibinfo{year}{2013}).
\newblock \urlprefix\url{https://link.aps.org/doi/10.1103/PhysRevB.88.161406}.

\bibitem{xu_manipulating_2017}
\bibinfo{author}{Xu, Y.} \& \bibinfo{author}{Jin, G.}
\newblock \bibinfo{title}{Manipulating topological inner-edge states in hybrid
  silicene nanoribbons}.
\newblock \emph{\bibinfo{journal}{Physical Review B}}
  \textbf{\bibinfo{volume}{95}}, \bibinfo{pages}{155425}
  (\bibinfo{year}{2017}).
\newblock \urlprefix\url{https://link.aps.org/doi/10.1103/PhysRevB.95.155425}.

\bibitem{yang_topological_2020}
\bibinfo{author}{Yang, J.-E.}, \bibinfo{author}{Lü, X.-L.},
  \bibinfo{author}{Zhang, C.-X.} \& \bibinfo{author}{Xie, H.}
\newblock \bibinfo{title}{Topological spin–valley filtering effects based on
  hybrid silicene-like nanoribbons}.
\newblock \emph{\bibinfo{journal}{New Journal of Physics}}
  \textbf{\bibinfo{volume}{22}}, \bibinfo{pages}{053034}
  (\bibinfo{year}{2020}).
\newblock
  \urlprefix\url{https://iopscience.iop.org/article/10.1088/1367-2630/ab84b4}.

\bibitem{qiao_spin-polarized_2011}
\bibinfo{author}{Qiao, Z.}, \bibinfo{author}{Yang, S.~A.},
  \bibinfo{author}{Wang, B.}, \bibinfo{author}{Yao, Y.} \&
  \bibinfo{author}{Niu, Q.}
\newblock \bibinfo{title}{Spin-polarized and valley helical edge modes in
  graphene nanoribbons}.
\newblock \emph{\bibinfo{journal}{Physical Review B}}
  \textbf{\bibinfo{volume}{84}}, \bibinfo{pages}{035431}
  (\bibinfo{year}{2011}).
\newblock \urlprefix\url{https://link.aps.org/doi/10.1103/PhysRevB.84.035431}.

\bibitem{jo_quantum_2021}
\bibinfo{author}{Jo, M.} \emph{et~al.}
\newblock \bibinfo{title}{Quantum {Hall} valley splitters and a tunable
  {Mach}-{Zehnder} interferometer in graphene}.
\newblock \emph{\bibinfo{journal}{Phys. Rev. Lett.}}
  \textbf{\bibinfo{volume}{126}}, \bibinfo{pages}{146803}
  (\bibinfo{year}{2021}).
\newblock
  \urlprefix\url{https://link.aps.org/doi/10.1103/PhysRevLett.126.146803}.

\bibitem{rickhaus_transport_2018}
\bibinfo{author}{Rickhaus, P.} \emph{et~al.}
\newblock \bibinfo{title}{Transport {Through} a {Network} of {Topological}
  {Channels} in {Twisted} {Bilayer} {Graphene}}.
\newblock \emph{\bibinfo{journal}{Nano Letters}} \textbf{\bibinfo{volume}{18}},
  \bibinfo{pages}{6725--6730} (\bibinfo{year}{2018}).
\newblock
  \urlprefix\url{https://pubs.acs.org/doi/10.1021/acs.nanolett.8b02387}.

\bibitem{wei_mach-zehnder_2017}
\bibinfo{author}{Wei, D.~S.} \emph{et~al.}
\newblock \bibinfo{title}{Mach-{Zehnder} interferometry using spin- and
  valley-polarized quantum {Hall} edge states in graphene}.
\newblock \emph{\bibinfo{journal}{Science Advances}}
  \textbf{\bibinfo{volume}{3}}, \bibinfo{pages}{e1700600}
  (\bibinfo{year}{2017}).
\newblock \urlprefix\url{https://www.science.org/doi/10.1126/sciadv.1700600}.

\bibitem{niyazov_coherent_2020}
\bibinfo{author}{Niyazov, R.~A.}, \bibinfo{author}{Aristov, D.~N.} \&
  \bibinfo{author}{Kachorovskii, V.~Y.}
\newblock \bibinfo{title}{Coherent spin transport through helical edge states
  of topological insulator}.
\newblock \emph{\bibinfo{journal}{npj Computational Materials}}
  \textbf{\bibinfo{volume}{6}}, \bibinfo{pages}{1--10} (\bibinfo{year}{2020}).
\newblock \urlprefix\url{https://www.nature.com/articles/s41524-020-00442-z}.

\bibitem{chen_quantum_2014}
\bibinfo{author}{Chen, W.}, \bibinfo{author}{Xue, Z.-Y.},
  \bibinfo{author}{Wang, Z.}, \bibinfo{author}{Shen, R.} \&
  \bibinfo{author}{Xing, D.~Y.}
\newblock \bibinfo{title}{Quantum computing through electron propagation in
  edge states of quantum spin {Hall} systems}.
\newblock \emph{\bibinfo{journal}{The European Physical Journal B}}
  \textbf{\bibinfo{volume}{87}}, \bibinfo{pages}{57} (\bibinfo{year}{2014}).
\newblock \urlprefix\url{https://doi.org/10.1140/epjb/e2014-40899-4}.

\bibitem{sundararaman2017jdftx}
\bibinfo{author}{Sundararaman, R.} \emph{et~al.}
\newblock \bibinfo{title}{{JDFTx}: Software for joint density-functional
  theory}.
\newblock \emph{\bibinfo{journal}{SoftwareX}} \textbf{\bibinfo{volume}{6}},
  \bibinfo{pages}{278--284} (\bibinfo{year}{2017}).
\newblock \urlprefix\url{https://doi.org/10.1016/j.softx.2017.10.006}.

\bibitem{hamann2013optimized}
\bibinfo{author}{Hamann, D.}
\newblock \bibinfo{title}{Optimized norm-conserving {Vanderbilt}
  pseudopotentials}.
\newblock \emph{\bibinfo{journal}{Physical Review B}}
  \textbf{\bibinfo{volume}{88}}, \bibinfo{pages}{085117}
  (\bibinfo{year}{2013}).
\newblock \urlprefix\url{https://doi.org/10.1103/PhysRevB.88.085117}.

\bibitem{van2018pseudodojo}
\bibinfo{author}{van Setten, M.~J.} \emph{et~al.}
\newblock \bibinfo{title}{The pseudodojo: Training and grading a 85 element
  optimized norm-conserving pseudopotential table}.
\newblock \emph{\bibinfo{journal}{Computer Physics Communications}}
  \textbf{\bibinfo{volume}{226}}, \bibinfo{pages}{39--54}
  (\bibinfo{year}{2018}).
\newblock \urlprefix\url{https://doi.org/10.1016/j.cpc.2018.01.012}.

\bibitem{pbe}
\bibinfo{author}{Perdew, J.~P.}, \bibinfo{author}{Burke, K.} \&
  \bibinfo{author}{Ernzerhof, M.}
\newblock \bibinfo{title}{Generalized gradient approximation made simple}.
\newblock \emph{\bibinfo{journal}{Physical Review Letters}}
  \textbf{\bibinfo{volume}{77}}, \bibinfo{pages}{3865} (\bibinfo{year}{1996}).
\newblock
  \urlprefix\url{https://journals.aps.org/prl/abstract/10.1103/PhysRevLett.77.3865}.

\bibitem{vasp1}
\bibinfo{author}{Kresse, G.} \& \bibinfo{author}{Furthm{\"u}ller, J.}
\newblock \bibinfo{title}{Efficient iterative schemes for ab initio
  total-energy calculations using a plane-wave basis set}.
\newblock \emph{\bibinfo{journal}{Physical review B}}
  \textbf{\bibinfo{volume}{54}}, \bibinfo{pages}{11169} (\bibinfo{year}{1996}).
\newblock
  \urlprefix\url{https://journals.aps.org/prb/abstract/10.1103/PhysRevB.54.11169}.

\bibitem{vasp2}
\bibinfo{author}{Kresse, G.} \& \bibinfo{author}{Furthm{\"u}ller, J.}
\newblock \bibinfo{title}{Efficiency of ab-initio total energy calculations for
  metals and semiconductors using a plane-wave basis set}.
\newblock \emph{\bibinfo{journal}{Computational materials science}}
  \textbf{\bibinfo{volume}{6}}, \bibinfo{pages}{15--50} (\bibinfo{year}{1996}).
\newblock
  \urlprefix\url{https://www.sciencedirect.com/science/article/pii/0927025696000080}.

\bibitem{vasp3}
\bibinfo{author}{Kresse, G.} \& \bibinfo{author}{Hafner, J.}
\newblock \bibinfo{title}{Ab initio molecular dynamics for liquid metals}.
\newblock \emph{\bibinfo{journal}{Physical review B}}
  \textbf{\bibinfo{volume}{47}}, \bibinfo{pages}{558} (\bibinfo{year}{1993}).
\newblock
  \urlprefix\url{https://journals.aps.org/prl/abstract/10.1103/PhysRevLett.77.3865}.

\bibitem{paw_pseudopotentials}
\bibinfo{author}{Kresse, G.} \& \bibinfo{author}{Joubert, D.}
\newblock \bibinfo{title}{From ultrasoft pseudopotentials to the projector
  augmented-wave method}.
\newblock \emph{\bibinfo{journal}{Physical Review B}}
  \textbf{\bibinfo{volume}{59}}, \bibinfo{pages}{1758} (\bibinfo{year}{1999}).
\newblock
  \urlprefix\url{https://journals.aps.org/prb/abstract/10.1103/PhysRevB.59.1758}.

\bibitem{ambrosetti2014long}
\bibinfo{author}{Ambrosetti, A.}, \bibinfo{author}{Reilly, A.~M.},
  \bibinfo{author}{DiStasio~Jr, R.~A.} \& \bibinfo{author}{Tkatchenko, A.}
\newblock \bibinfo{title}{Long-range correlation energy calculated from coupled
  atomic response functions}.
\newblock \emph{\bibinfo{journal}{The Journal of chemical physics}}
  \textbf{\bibinfo{volume}{140}}, \bibinfo{pages}{18A508}
  (\bibinfo{year}{2014}).
\newblock \urlprefix\url{https://doi.org/10.1063/1.4865104}.

\bibitem{pawpyseed}
\bibinfo{author}{Bystrom, K.}, \bibinfo{author}{Broberg, D.},
  \bibinfo{author}{Dwaraknath, S.}, \bibinfo{author}{Persson, K.~A.} \&
  \bibinfo{author}{Asta, M.}
\newblock \bibinfo{title}{Pawpyseed: Perturbation-extrapolation band shifting
  corrections for point defect calculations}.
\newblock \emph{\bibinfo{journal}{arXiv preprint arXiv:1904.11572}}
  (\bibinfo{year}{2019}).
\newblock \urlprefix\url{https://arxiv.org/abs/1904.11572}.

\bibitem{fang2015quantum}
\bibinfo{author}{Fang, Y.} \emph{et~al.}
\newblock \bibinfo{title}{Quantum spin {Hall} states in stanene/{Ge} (111)}.
\newblock \emph{\bibinfo{journal}{Scientific Reports}}
  \textbf{\bibinfo{volume}{5}}, \bibinfo{pages}{1--8} (\bibinfo{year}{2015}).
\newblock \urlprefix\url{https://www.nature.com/articles/srep14196}.

\bibitem{InSb_substrate}
\bibinfo{author}{Xu, C.-Z.} \emph{et~al.}
\newblock \bibinfo{title}{Gapped electronic structure of epitaxial stanene on
  insb (111)}.
\newblock \emph{\bibinfo{journal}{Physical Review B}}
  \textbf{\bibinfo{volume}{97}}, \bibinfo{pages}{035122}
  (\bibinfo{year}{2018}).
\newblock \urlprefix\url{https://link.aps.org/doi/10.1103/PhysRevB.97.035122}.

\bibitem{hbn_substrate}
\bibinfo{author}{Wang, D.}, \bibinfo{author}{Chen, L.}, \bibinfo{author}{Wang,
  X.}, \bibinfo{author}{Cui, G.} \& \bibinfo{author}{Zhang, P.}
\newblock \bibinfo{title}{The effect of substrate and external strain on
  electronic structures of stanene film}.
\newblock \emph{\bibinfo{journal}{Physical Chemistry Chemical Physics}}
  \textbf{\bibinfo{volume}{17}}, \bibinfo{pages}{26979--26987}
  (\bibinfo{year}{2015}).
\newblock \urlprefix\url{https://doi.org/10.1039/C5CP04322K}.

\end{thebibliography}
    
\section*{Acknowledgements}
    We thank the US Department of Energy for support under grant DE-SC0020128. J.C. acknowledges support from the Department of Energy Computational Science Graduate Fellowship (DOE CSGF) under Award Number DE-FG02-97ER25308. J.C. thanks Kyle Bystrom for assistance with the pawpyseed code as well as David Lim and Cameron Owen for useful discussion.
    
\section*{Author contributions}
    J.C. performed the first-principles simulations. J.C. and O.D. prepared the figures. O.D. and M.R.H. conceived the decoration scheme, performed theoretical calculations, and guided the first-principles calculations. T.H. and B.K. supervised all aspects of the project. All authors jointly wrote the manuscript.
    
\section*{Data availability}

The data that support the findings of this study are available from the corresponding author upon reasonable request.

\section*{Code availability}

The codes that support the findings of this study are available from the corresponding author upon reasonable request.

\newpage

\newcommand{\beginsupplement}{%
        \setcounter{table}{0}
        \renewcommand{\thetable}{S\arabic{table}}%
        \setcounter{figure}{0}
        \renewcommand{\thefigure}{S\arabic{figure}}%
        \setcounter{equation}{0}
        \renewcommand{\theequation}{S\arabic{equation}}
        \setcounter{section}{0}
        \renewcommand{\thesection}{S\arabic{section}}%
}

\beginsupplement

\begin{center}
{ \Large \textbf{Supplementary Material:}}
{ \Large \textbf{Engineering ideal helical topological networks in stanene via Zn decoration}}

\vspace{12pt}
\vspace{12pt}

{ \large {Jennifer Coulter}$^1$,
{Mark R. Hirsbrunner}$^2$,
{Oleg Dubinkin}$^2$,
{Taylor L. Hughes}$^2$,
and {Boris Kozinsky}}$^1$ \\
\vspace{12pt}
    {\footnotesize{$^1$Harvard John A. Paulson School of Engineering and Applied Sciences, Harvard University, Cambridge, MA, 02138, USA}} \\
    {\footnotesize{$^2$Department of Physics and Institute for Condensed Matter Theory, University of Illinois at Urbana-Champaign, Urbana, IL 61801, USA}} \\
\end{center}


\begin{center}
{ \large \textbf{Buckling Height Considerations}}
\end{center}

\normalsize 
In our calculations we fix stanene to its relaxed ground state structure as determined using DFT, but realistically, the structure of stanene can change based on its substrate as a result of lattice constant matching. Specifically, biaxial strain affects the buckling height and therefore the band gap of stanene. Here we calculate the band gap for bare stanene and the V and T decorated structures as a function of buckling height (as this is what is often reported in experimental studies of stanene) to ensure that our conclusions remain valid for stanene on substrates, rather than only for free standing stanene. 

\begin{figure}
    \centering
    \includegraphics[width=1.0\textwidth]{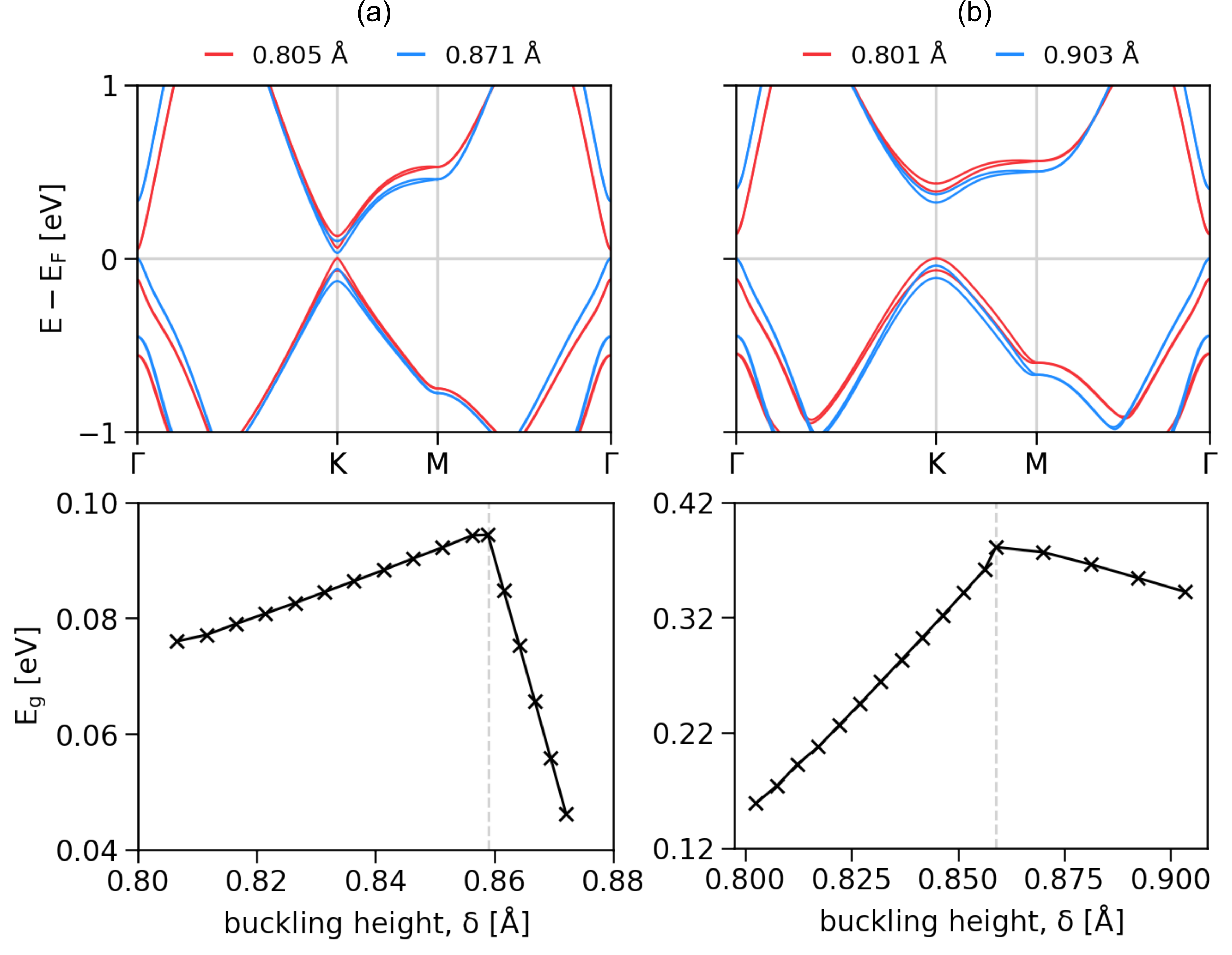}
    \caption{
    \textbf{Biaxially strained stanene to simulate substrate effects.}
    The top row shows the electronic structure of (a) V- and (b) T-decorated stanene, with the maximum tested strain value shown in red and the minimum value shown in blue. The bottom row plots band gap values for buckling heights extracted from the strained stanene calculations, with the unstrained buckling height indicated by the dashed vertical lines.
    }
    \label{fig:strain_gaps}
\end{figure}

We calculated band gap values for a range of buckling heights from $\sim$0.8~\AA~to $\sim$0.9~\AA, which covers a variety of previously used substrates \cite{fang2015quantum, InSb_substrate, hbn_substrate}. Fig.~\ref{fig:strain_gaps}a and b show the electronic structures of V- and T-decorated stanene for the maximum (red) and minimum (blue) strain values tested. From these plots it is clear that varying the buckling height shifts the bands near the Fermi level, but does not strongly affect the bands otherwise. 

For V-decorated stanene we find the structure remains insulating for buckling heights from 0.805~\AA~to 0.871~\AA. Outside this range, the bands at the $\Gamma$ point overlap the Fermi energy and the structure becomes metallic. The T-decorated structure features a larger band gap, and therefore is relatively impervious to changes in buckling height. We found T-decorated stanene to be insulating for the entire range of tested strains. Because both the V and T decorated adatom structures remained insulating for a reasonably wide range of strain values, the conclusions drawn in this work will apply to stanene on most insulating substrates with a reasonably good lattice match.

\begin{center}
\large{\textbf{Alternate Candidate Adatoms}}
\end{center}

\normalsize
In addition to Zn adatoms, we identified four other candidate adatom decorations for inducing the QVH phase in stanene. Decorating the T sublattice with Be adatoms results in a moderately doped metallic phase with massive Dirac cones shifted away from the $\v{K}$ and $\v{K}'$ points. The shift of the Dirac cones is of no consequence for QVH-QSH interfaces, as the spin Chern number will change across the interface regardless. Decoration of stanene with the other candidate adatoms (Li, Na, K, Ca, Sc) leaves the Dirac cones at the $\v{K}$ and $\v{K}'$ points, but result in significant electron doping leaving the Fermi level deep in the valence or conduction bands. This can be remedied by a dual-gate setup that can change the electron density without applying an out-of-plane electric field that would otherwise interfere with the formation of the QVH phase. These candidates are enumerated in Table~\ref{tab:adatoms}, along with the resulting valley Chern number, band gap, diffusion barrier, operating temperature, and energy shifts required to place the Fermi level in the middle of the topological band gap. Some of these atoms have significantly larger diffusion barriers than Zn, but unfortunately the requirement of gating these systems eliminates the advantages of the adatom decoration scheme because device geometries would be limited by more restrictive gate fabrication technology.

\begin{table}
    \centering
    \begin{tabular}{|c | c | c | c | c | c | c | c |} 
         \hline
         Adatom & Site & $C_v$ & Gap (eV) & Diffusion Barrier (meV) & Fermi Level Shift (eV) \\
         \hline\hline
         Be$^*$ & T & N/A & 0.4236 & 12.0070 ($\sim139$ K) & 0.2779  \\
         \hline
         K & V & $+2$ & 0.1323 & 33.0151 ($\sim383$ K) & -0.5577  \\
         \hline
         Ca & V & $+2$ & 0.0886 & 322.4253 ($>400$ K) & -0.8228  \\
         \hline
         Sc & V & $+2$ & 0.2055 & 645.7683 ($>400$ K) & -0.9683 \\
         \hline
    \end{tabular}
    \caption{The valley Chern number, band gap, diffusion barrier (and corresponding operating temperature), and Fermi level shift required for the four other candidate decoration strategies. The Fermi level shift here indicates the energy difference required to shift to the chemical potential to the middle of the topological gap. The valley Chern number of the Be decoration is defined with respect to a different vector in momentum space than the other decoration strategies so to avoid confusion, we do not assign it a value.}
    \label{tab:adatoms}
\end{table}

\begin{center}
\large{\textbf{Finite Size Effects}}
\end{center}

\begin{figure}
    \centering
    \includegraphics[width=1.0\textwidth]{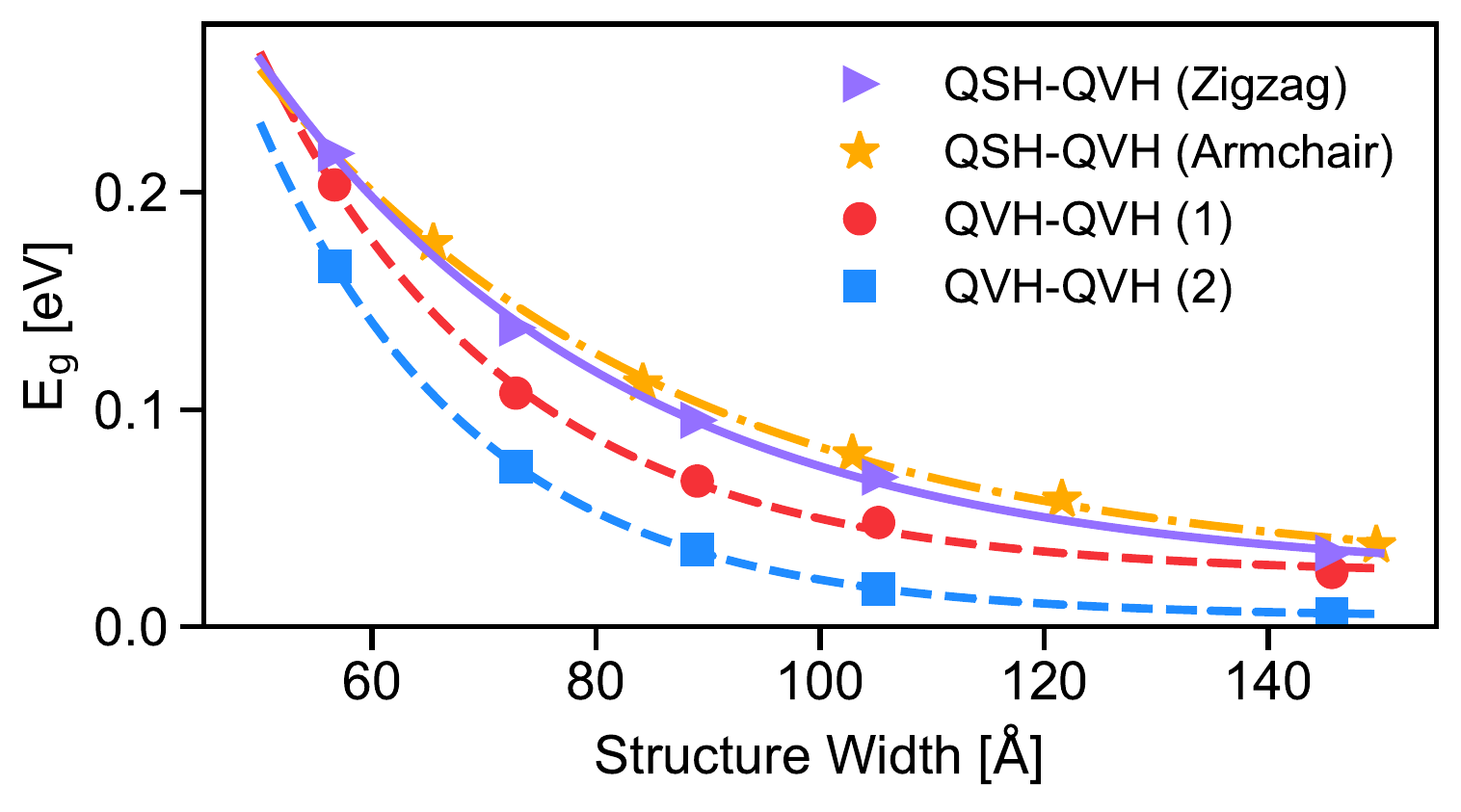}
    \caption{
    \textbf{Width dependence of the interface spectrum gaps.}
    The gaps of the interface spectra for each ribbon type as a function of system width. For each ribbon, the width of the T-decorated region is held constant at 10 unit cells while the widths of the V-decorated and bare regions are varied. Each gap decays exponentially to some finite value, the origin of which is wavefunction overlap in the T-decorated regions.
    }
    \label{fig:finite_size}
\end{figure}

\normalsize
The interface states of our first-principles calculations exhibit small gaps that arise from wavefunction overlap caused by the finite size of the system. These gaps spoil the technological advantages of the interface states, so it is important to confirm that they will vanish when the interfaces are sufficiently well separated. To this end, we plot the size of the interface spectrum gaps in each ribbon as a function of the width in Fig.~\ref{fig:finite_size}. For all ribbons, the width of the T-decorated regions are fixed at 10 unit cells and the widths of the bare and V-decorated regions are varied. The QVH-QVH ribbon has two gaps because each interface hosts two pairs of valley-momentum locked states. These gaps are well-fit by exponential functions of the form $E_g(x)=E_0e^{-x/\lambda}+E_T$. We report the values of these fit parameters in Table~\ref{tab:gap_fit}. The fitting parameter $E_0$ gives the zero-width limit of the interface gap, but since there is no interface when the width is zero, this parameter has no physical significance. The decay length $\lambda$ determines how rapidly the interface gap decreases as the width is increased, and the values obtained here closely match those obtained in the main text from the ratio of the Fermi velocity to the bulk gap. The long-width limit, $E_T$, of the interface gap is caused by wavefunction overlap in the constant-width T-decorated region. The bulk gap in the T-decorated region is much larger than the bare or V-decorated regions, so this contribution to the interface gap will vanish with a small increase in the size of the T-decorated regions. This analysis confirms that the interface spectrum gaps are a finite size effect.

\begin{table}
    \centering
    \begin{tabular}{| c | c | c | c |} 
         \hline
         Ribbon & $E_0$ (eV) & $\lambda$ (\textrm{\AA}) & $E_T$ (eV) \\
         \hline\hline
         QSH-QVH (zigzag) & 1.12 & 32.36 & 0.023 \\
         \hline
         QSH-QVH (armchair) & 0.92 & 36.19 & 0.024 \\
         \hline
         QVH-QVH (1) & 2.26 & 22.33 & 0.024 \\
         \hline
         QVH-QVH (2) & 3.03 & 19.32 & 0.0046 \\
         \hline
    \end{tabular}
    \caption{The exponential fit parameters for each interface spectrum gap, $E_g(x)=E_0e^{-x/\lambda}+E_T$. The fitting parameter $E_0$ has no physical significance as there is no concept of an interface spectrum in the zero-width limit. $\lambda$ is the decay length of the gap with respect to system width and $E_T$ is the finite long-width limit of the interface gap caused by wavefunction overlap in the T-decorated region.}
    \label{tab:gap_fit}
\end{table}

\end{document}